\documentclass[journal]{IEEEtran}
\usepackage{amsmath,amsfonts}
\usepackage{graphicx}
\usepackage{grffile}
\usepackage{cite}
\usepackage{booktabs}
\usepackage{url}
\usepackage{stfloats}
\usepackage[caption=false,font=footnotesize]{subfig}
\usepackage[ruled,vlined]{algorithm2e}
\usepackage{tabularx}
\usepackage[table]{xcolor}
\usepackage{array}
\usepackage{makecell}
\begin{document}

\title{Real-Time Surrogate Modeling for Fast Transient Prediction in Inverter-Based Microgrids Using CNN and LightGBM}

\author{Osasumwen Cedric Ogiesoba-Eguakun,~\IEEEmembership{Member,~IEEE,}, Kaveh Ashenayi, ~\IEEEmembership{Senior Member,~IEEE}, \\ Suman Rath,~\IEEEmembership{Member,~IEEE}
\thanks{This work was conducted as part of the graduate research activities at the University of Tulsa.\\ 
(Corresponding author: Osasumwen Cedric Ogiesoba-Eguakun.)
 
O. C. Ogiesoba-Eguakun, K. Ashenayi, and S. Rath are with the Department of Electrical and Computer Engineering, The University of Tulsa, Tulsa, OK 74104, USA (e-mail: oco1411@utulsa.edu, kash@utulsa.edu, suman-rath@utulsa.edu).}

}



\maketitle

\begin{abstract}
Real-time monitoring of inverter-based microgrids is essential for stability, fault response, and operational decision-making. However, electromagnetic transient (EMT) simulations, required to capture fast inverter dynamics, are computationally intensive and unsuitable for real-time applications. This paper presents a data-driven surrogate modeling framework for fast prediction of microgrid behavior using convolutional neural networks (CNN) and Light Gradient Boosting Machine (LightGBM). The models are trained on a high-fidelity EMT digital twin dataset of a microgrid with ten distributed generators under eleven operating and disturbance scenarios, including faults, noise, and communication delays. A sliding-window method is applied to predict important system variables, including voltage magnitude, frequency, total active power, and voltage dip. The results show that model performance changes depending on the type of variable being predicted. The CNN demonstrates high accuracy for time-dependent signals such as voltage, with an $R^2$ value of 0.84, whereas LightGBM shows better performance for structured and disturbance-related variables, achieving an $R^2$ of 0.999 for frequency and 0.75 for voltage dip. A combined CNN+LightGBM model delivers stable performance across all variables. Beyond accuracy, the surrogate models also provide major improvements in computational efficiency. LightGBM achieves more than $1000\times$ speedup and runs faster than real time, while the hybrid model achieves over $500\times$ speedup with near real-time performance. These findings show that data-driven surrogate models can effectively represent microgrid dynamics. They also support real-time and faster-than-real-time predictions. As a result, they are well-suited for applications such as monitoring, fault analysis, and control in inverter-based power systems.
\end{abstract}

\begin{IEEEkeywords}
Surrogate modeling, microgrid, inverter-based microgrids, real-time simulation, convolutional neural networks, LightGBM, electromagnetic transient simulation, distributed energy resources.
\end{IEEEkeywords}

\section{Introduction}
\IEEEPARstart{T}{he} increasing use of distributed energy resources, such as solar panels, batteries, and converter-based generators, is moving power systems toward inverter-based structures \cite{samal2025review,aghdam2025navigating}. These technologies improve flexibility and support clean energy, but they also introduce faster system dynamics due to power electronic control loops \cite{aghdam2025navigating,thwe2025digital}. Because of this, accurate analysis of inverter-based microgrids requires electromagnetic transient (EMT) simulations that can capture very fast changes, control interactions, and system responses to disturbances. EMT simulations provide high accuracy, but they are computationally heavy and not suitable for real-time use \cite{mohammadi2024surrogate,marrel2024probabilistic}. In real microgrid operation, stability monitoring, disturbance detection, and emergency response all need fast prediction of system behavior. This leads to a strong need for efficient surrogate models that can capture system dynamics while keeping key physical properties. Data-driven surrogate modeling is becoming a strong solution for speeding up power system simulations \cite{mohammadi2024surrogate,prina2024machine}. Machine learning models can learn how measurements relate to system behavior, making it possible to predict quickly without solving complex equations, especially with the availability of open power system datasets and simulation frameworks \cite{aravena2025open}. Earlier work has used tree-based models, neural networks, and reduced-order models \cite{islam2025machine, lim2025power, zhang2021critical}. However, many of these studies use low-resolution data, limited disturbance cases, or steady-state assumptions, which makes it hard to capture fast inverter dynamics and transient behavior \cite{oelhaf2025scoping,barreto2025cyber}. Also, there is still a limited understanding of how different machine learning models perform for different types of microgrid variables, especially under disturbances and uncertainty.

To address these issues, this paper develops and evaluates surrogate models for real-time prediction of inverter-based microgrid dynamics using convolutional neural networks (CNN) and Light Gradient Boosting Machine (LightGBM). The models are trained using a high-quality digital twin dataset generated from an EMT simulation of a microgrid with ten distributed generation units \cite{ogiesoba2026high}. The dataset includes synchronized multi-channel measurements and eleven operating and disturbance scenarios. These scenarios include electrical events such as load changes, faults, and generator trips, as well as cyber-physical effects like measurement noise and communication delays \cite{ogiesoba2026high,barreto2025cyber}. A sliding-window learning approach is used to capture how the system changes over time. The surrogate models are designed to predict important microgrid variables, including voltage magnitude, system frequency, total active power, and voltage dip severity. Their performance is assessed based on accuracy and also under out-of-distribution conditions, where noise and delayed signals are used to simulate real-world uncertainty. The results show that how well the model performs depends on the variable being predicted. CNN models are effective at learning time-dependent waveform patterns and can remain robust even under degraded input conditions, especially for voltage prediction \cite{shi2020convolutional,ogiesoba5264265robust}. LightGBM models perform better for aggregated system variables and disturbance-related values. This confirms that no single model performs best across all prediction tasks in microgrid systems. In addition to accuracy, the surrogate models are much faster than EMT simulations. They achieve very large speed improvements and can support real-time or even faster-than-real-time prediction.

The main contributions of this paper are as follows:
\begin{itemize}
    \item Development of surrogate modeling frameworks using CNN and LightGBM for predicting dynamic behavior in inverter-based microgrids

    \item Evaluation of model performance across different system variables, showing how model choice depends on the physical nature of the variable

    \item Testing of model robustness under disturbances and uncertainty, including measurement noise and communication delays

    \item Detailed runtime comparison showing large speed improvements compared to EMT simulation and confirming real-time capability

\end{itemize}

The rest of the paper is organized as follows. Section II reviews related work in surrogate modeling and data-driven power system analysis. Section III describes the dataset and problem setup. Section IV describes the surrogate modeling method. Section V shows the results, including accuracy, robustness, and runtime performance. Section VI concludes the paper and highlights future work.

\section{Related Work}

Surrogate modeling and data-driven approaches are getting more attention in power systems because they help reduce the heavy computation required for high-detail simulations \cite{aghazadeh2024digital,kabir2024digital,mbasso2025digital}. As inverter-based microgrids become more complex, there is a growing need for fast models that can predict system behavior for real-time monitoring and control. Early research in surrogate modeling used reduced-order and linear models to simplify system behavior \cite{mohammadi2024surrogate,marrel2024probabilistic,prina2024machine}. These methods improve speed, but they often cannot capture the nonlinear behavior caused by power electronic converters and control interactions in modern microgrids. Because of this, machine learning methods have become a strong alternative, since they can learn complex nonlinear relationships directly from data. Tree-based models like gradient boosting and random forests are commonly used in power systems \cite{yang2024research,fu2025small,islam2025machine}. They work well with structured data and can model nonlinear patterns. These models have been applied to load forecasting, fault detection, and stability analysis \cite{r2024machine,oelhaf2025scoping}. However, they rely on manual feature design and may have difficulty capturing time-based behavior.

Deep learning uses CNNs and recurrent neural networks (RNNs) to model time-series behavior in power systems \cite{zhang2021critical,shi2020convolutional,lee2023power}. CNN models can extract patterns from high-frequency data, while LSTM networks focus on learning time and sequence relationships. These methods work well for modeling transient behavior and waveform features in inverter-based systems. However, deep learning models need more data, more tuning, and more computation than tree-based models. Recently, neural networks and hybrid methods have been used to study surrogate modeling for dynamic power system simulation \cite{cheng2024machine,ellinas2025physics,du2025development}. These approaches aim to learn the input–output behavior of detailed simulation models so that system responses can be predicted faster under different conditions. However, many existing studies have some limitations. These issues include using low-resolution or steady-state data, studying only a few disturbance scenarios, not testing under uncertainty like noise or communication delays, and not comparing runtime with EMT simulations in detail \cite{aygul2024benchmark,oelhaf2025scoping}. Fast transient behavior in inverter-based microgrids is still difficult to capture. The combined effects of inverter control, network conditions, and disturbances create nonlinear and time-dependent behavior that simple models cannot easily represent. Although machine learning has been applied to tasks such as transient security assessment using deep learning models \cite{pournabi2022power}, there is still a limited understanding of how different models perform for variables such as voltage, frequency, and power.

This paper addresses these gaps by developing and evaluating surrogate models using both CNN and LightGBM in a single framework. This work goes beyond earlier studies by using a high-quality EMT-based digital twin dataset with multiple disturbance scenarios \cite{ogiesoba2026high}. It evaluates model performance under out-of-distribution conditions, such as noise and communication delays \cite{aygul2024benchmark,zhang2025frequency}. It also compares how different models perform across various prediction variables, linking model choice to the physical properties of each variable. In addition, runtime analysis shows that the models can be used for real-time or faster-than-real-time microgrid analysis. Table~\ref{tab:related_work_comparison} summarizes representative recent studies and highlights how the present work differs in terms of data fidelity, disturbance diversity, out-of-distribution testing, and real-time benchmarking.

\begin{table*}[!ht]
\centering
\caption{Comparison of Representative Recent Studies on Data-Driven and Surrogate Modeling Approaches for Power-System Dynamics}
\label{tab:related_work_comparison}
\renewcommand{\arraystretch}{1.15}
\setlength{\tabcolsep}{4pt}
\footnotesize
\begin{tabular*}{\textwidth}{@{\extracolsep{\fill}} l l l l l l p{4.2cm}}
\toprule
\rowcolor{gray!15}
\textbf{Reference} & \textbf{System / Focus} & \textbf{Model Type} &
\textbf{\makecell[l]{EMT High-\\Frequency Data}} &
\textbf{\makecell[l]{Multiple\\Disturbances}} &
\textbf{\makecell[l]{OOD /\\Robustness Test}} &
\textbf{Main Limitation / Scope} \\
\midrule

\rowcolor{gray!8}
\cite{shi2020convolutional} & Transient stability & CNN & No & Limited & No & Focused on stability classification, not multi-target surrogate regression \\

\rowcolor{white}
\cite{pournabi2022power} & \makecell[l]{Transient security\\assessment} & Deep learning & No & Limited & Partial & Emphasis on security assessment under partial observability \\

\rowcolor{gray!8}
\cite{lee2023power} & Transient stability & CNN + saliency map & No & Limited & No & Primarily classification and interpretability oriented \\

\rowcolor{white}
\cite{bogodorova2024fast} & Small-signal stability & Deep CNN & No & Limited & No & Focused on small-signal stability, not EMT waveform prediction \\

\rowcolor{gray!8}
\cite{li2024power} & \makecell[l]{Voltage vulnerability\\assessment} & CNN & No & Limited & No & Application-specific vulnerability assessment \\

\rowcolor{white}
\cite{yang2024research} & \makecell[l]{Small-signal\\stability correction} & LightGBM & No & Limited & No & Structured stability features, not EMT multi-channel dynamics \\

\rowcolor{gray!8}
\cite{fu2025small} & \makecell[l]{Wind-power\\stability prediction} & LightGBM & No & Limited & Partial & Focused on wind-power small-signal stability \\

\rowcolor{white}
\cite{mohammadi2024surrogate} & \makecell[l]{OPF surrogate\\modeling review} & Surrogate models & No & N/A & N/A & Steady-state OPF surrogates, not dynamic EMT microgrid prediction \\

\rowcolor{gray!8}
\cite{ogiesoba2026high} & \makecell[l]{Inverter-based\\microgrid dataset} & Digital twin dataset & Yes & Yes & Yes & Dataset generation paper; does not train surrogate predictors \\

\rowcolor{white}
This work & \makecell[l]{Inverter-based\\microgrid\\dynamic prediction} & CNN + LightGBM & \textbf{Yes} & \textbf{Yes} & \textbf{Yes} & \textbf{Real-time multi-target surrogate regression with runtime benchmarking} \\

\bottomrule
\end{tabular*}
\end{table*}

\section{System Description and Problem Formulation}

\subsection{Inverter-Based Microgrid Model}

\begin{figure}[ht]
\centering
\includegraphics[width=\columnwidth]{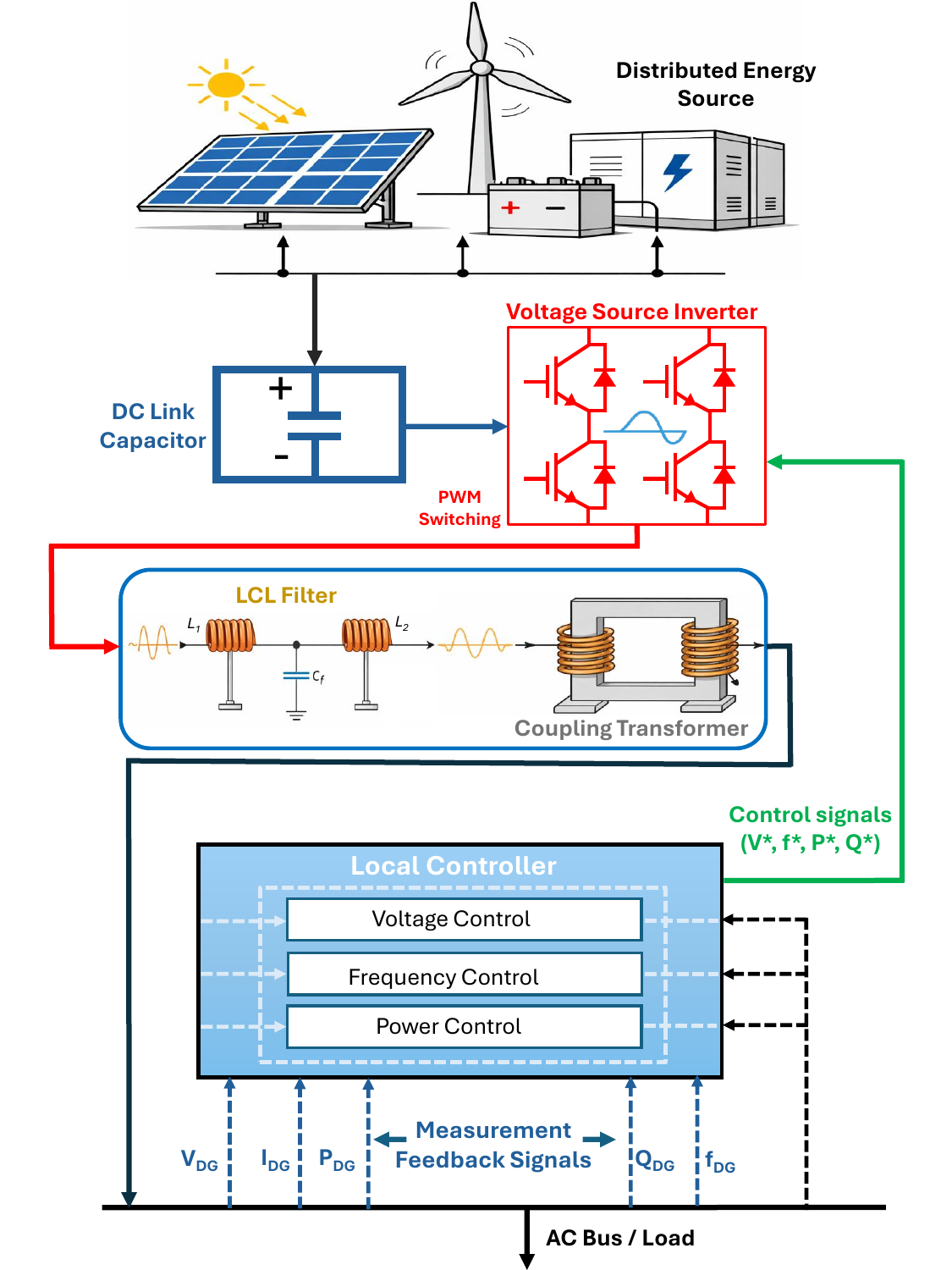}
\caption{Internal structure of a representative distributed generation (DG) unit in the inverter-based microgrid. Each DG unit consists of a distributed energy source, a DC-link capacitor, and a voltage source inverter (VSI) that operates using PWM switching. The inverter output is filtered by an LCL filter and connected to the AC bus via a coupling transformer. A local controller regulates voltage, frequency, and power using control signals $(V^*, f^*, P^*, Q^*)$, based on measurement feedback $(V_{DG}, I_{DG}, P_{DG}, Q_{DG}, f_{DG})$. The same structure is assumed for all DG units.}
\label{fig:dg_structure}
\end{figure}

This study employs a high-fidelity digital twin model of an inverter-based microgrid, developed in MATLAB/Simulink using electromagnetic transient (EMT) simulation \cite{ogiesoba2026high,jiang2024digital,von2023power}. The microgrid has ten distributed generation (DG) units connected through power electronic inverters and operates in a grid-connected mode. Each DG unit has local control systems that regulate voltage, frequency, and power output, allowing coordinated operation across the network. Fig.~\ref{fig:dg_structure} shows the internal structure of a representative DG unit used in this study.

Each DG unit is built on a standard inverter-based structure, where a power electronic converter links the energy source to the AC network. The DC-link capacitor maintains a stable voltage, and the inverter produces AC output through high-frequency switching. An LCL filter is used to reduce unwanted harmonics, while a coupling transformer links the unit to the AC bus. A local controller regulates voltage, frequency, and power using real-time measurements. This setup allows quick system response and accurate modeling of inverter behavior in EMT simulations. The key electrical and control parameters of the microgrid are summarized in Table~\ref{tab:microgrid_params}. The microgrid consists of multiple feeders, loads, and connection points, allowing realistic operation under varying conditions. Since the system is inverter-dominated, it shows fast transient responses due to power electronic control loops \cite{samal2025review,thwe2025digital}. For this reason, EMT-level simulation is needed to accurately capture system dynamics. The simulation uses a microsecond time resolution to capture high-frequency effects and control interactions.

\begin{table}[t]
\centering
\caption{Key Parameters of the Inverter-Based Microgrid}
\label{tab:microgrid_params}
\begin{tabular}{lc}
\toprule
\textbf{Parameter} & \textbf{Value} \\
\midrule
Filter resistance, $R_f$ & $0.1~\Omega$ \\
Filter inductance, $L_f$ & $4~\mathrm{mH}$ \\
Filter capacitance, $C_f$ & $200~\mu\mathrm{F}$ \\
Line resistance, $R_c$ & $0.1~\Omega$ \\
Line parameters (2, 4, 6, 8) & $1.5~\mathrm{mH} + 0.1~\Omega$ \\
Line parameters (1, 3, 5, 7, 9) & $0.5~\mathrm{mH} + 0.07~\Omega$ \\
Switching frequency, $f_{sw}$ & $10~\mathrm{kHz}$ \\
DG rating, $S$ & $10~\mathrm{kVA}$ \\
Active power droop coefficient, $m_P$ & $1\times10^{-4}$ \\
Reactive power droop coefficient, $n_Q$ & $1\times10^{-4}$ \\
DC-link voltage, $V_{dc}$ & $1000~\mathrm{V}$ \\
Nominal frequency, $f_n$ & $60~\mathrm{Hz}$ \\
Nominal angular frequency, $\omega_n$ & $2\pi \times 60~\mathrm{rad/s}$ \\
\bottomrule
\end{tabular}
\end{table}

\subsection{Dataset Generation and Disturbance Scenarios}

A large dataset is generated from the digital twin model by simulating different operating conditions and disturbances. It includes synchronized multi-channel measurements from all DG units and system buses \cite{ogiesoba2026high}. Each sample has 38 features, such as three-phase voltages and currents, active and reactive power, and frequency measurements from each DG unit. To ensure the dataset is realistic and diverse, eleven operating scenarios are included: \textbf{Normal operation, Load step changes, Voltage sag events, Ramp disturbances, Frequency ramp variations, Generator trip events, Tie-line disconnection, Reactive power disturbances, Single-line-to-ground faults, Measurement noise injection, and Communication delay effects.}

\begin{figure}[ht]
\centering
\includegraphics[width=\columnwidth]{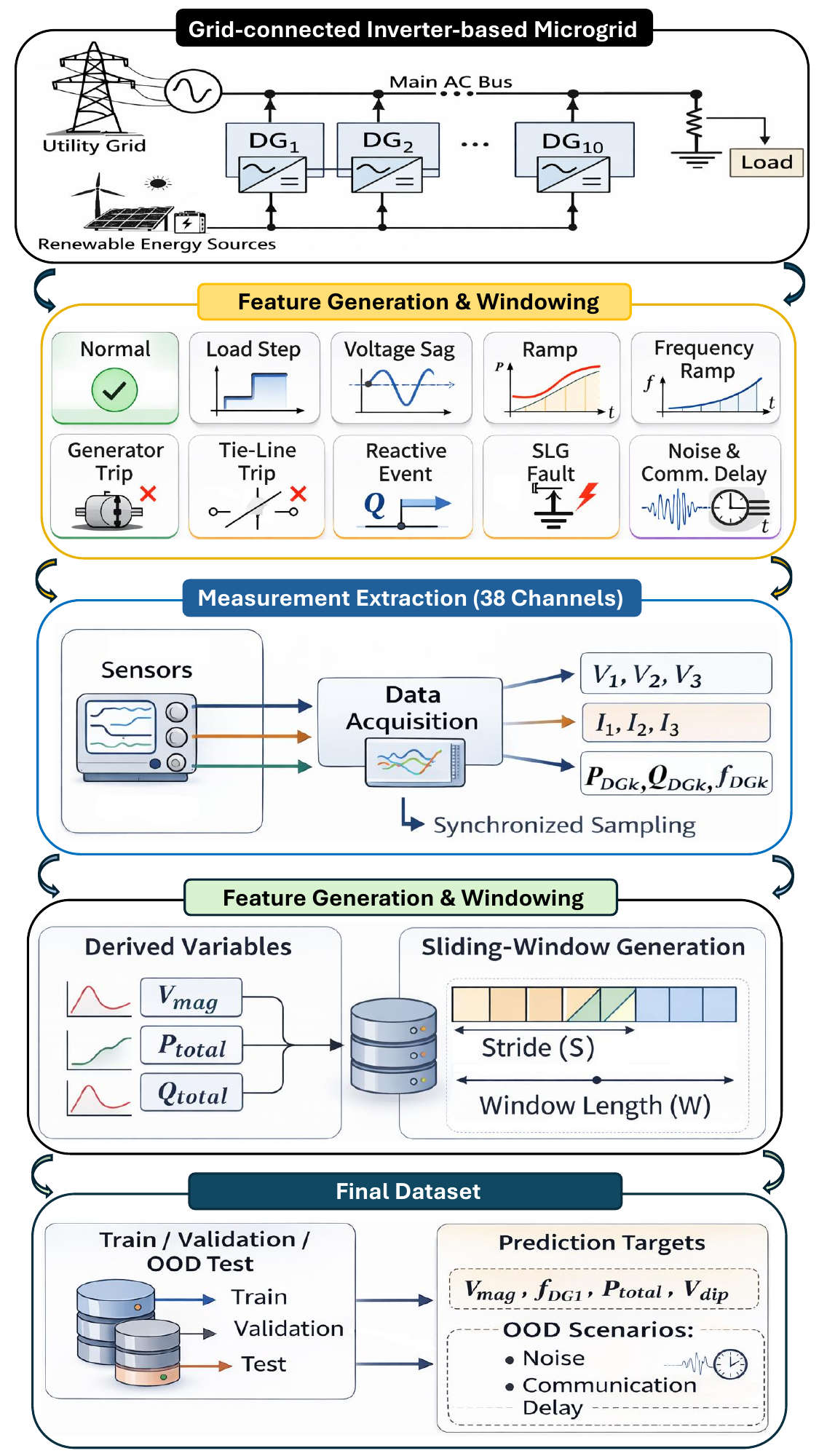}
\caption{EMT digital twin microgrid and dataset generation framework. A grid-connected inverter-based microgrid is simulated under multiple disturbance scenarios. Synchronized measurements are collected and processed to extract key electrical variables $(V_1,V_2,V_3, I_1,I_2,I_3, P_{DG_k}, Q_{DG_k}, f_{DG_k})$. Derived features $(V_{mag}, P_{total}, Q_{total})$ are computed and segmented using a sliding-window approach with window length $W$ and stride $S$. The resulting dataset is organized into training, validation, and OOD test sets for surrogate modeling.}
\label{fig:framework}
\end{figure}

These scenarios create more realistic testing conditions by integrating both physical and cyber-physical effects \cite{barreto2025cyber,thwe2025digital}. The dataset is split into training, validation, and testing sets. The test set includes out-of-distribution (OOD) cases, which are noise and communication delays datasets. These cases are used to assess model robustness. Figure~\ref{fig:framework} shows the EMT digital twin and dataset generation workflow. After simulation, the collected data undergoes feature extraction and sliding-window segmentation to create structured time-series inputs for modeling and evaluation. Phase angle measurements were not explicitly included, as these variables capture the dominant dynamics of inverter-based microgrids, including system stability, power balance, and control behavior. Moreover, phase angle information is implicitly reflected, making explicit inclusion unnecessary for the considered prediction tasks. Sliding-window inputs and engineered features are then used for multi-target prediction.

\subsection{Feature Engineering and Input Representation}

To improve model learning and capture important system behavior, additional features are created from the raw data. For example, voltage magnitude is calculated from three-phase voltages, and total active and reactive power are computed by combining values from all DG units.

\[
P_{\text{total}} = \sum_{k=1}^{N_{DG}} P_{DG_k}, \quad
Q_{\text{total}} = \sum_{k=1}^{N_{DG}} Q_{DG_k}
\]

where $N_{DG}$ denotes the total number of distributed generator units. $P_{DG_k}$ and $Q_{DG_k}$ are the active and reactive power outputs of DG unit $k$, respectively.

A sliding-window method is used to include time information in the model. For each prediction, a fixed-length window of past data is used as input.

\[
X_t = \{x_{t-W+1}, x_{t-W+2}, \ldots, x_t\}
\]

where $x_t \in \mathbb{R}^{d}$ denotes the input measurement vector at time step $t$.

Statistical features such as the mean, standard deviation, minimum, maximum, and the last value in each window are extracted for tree-based models. In contrast, sequence-based models use the raw time-series data within the window directly.

\subsection{Problem Formulation}

By learning the relationship between recent measurements and future system states, this study develops surrogate models to predict microgrid dynamics.

Let
\[
X_t \in \mathbb{R}^{W \times d}
\]
represent a sequence of measurements over a time window of length $W$, where $d$ is the number of input features. The surrogate modeling problem is to learn a nonlinear mapping:
\[
f_{\theta}: \mathbb{R}^{W \times d} \rightarrow \mathbb{R}^{m}
\]
$m$ refers to the number of prediction targets, and $\theta$ represents the trainable parameters of the surrogate model.

This mapping is defined as:
\[
y_t = f_{\theta}(X_t)
\]
where $y_t \in \mathbb{R}^{m}$ is the vector of target variables.

The learning objective is to minimize the prediction error across $N$ samples:
\[
\mathcal{L}(\theta) = \frac{1}{N} \sum_{t=1}^{N} \| y_t - \hat{y}_t \|_2^2
\]

where $\hat{y}_t = f_{\theta}(X_t)$ refers to the predicted output and $\|\cdot\|_2$ denotes the Euclidean norm.

The models are designed to predict key variables, which include voltage magnitude to represent voltage behavior, frequency of a DG unit to indicate system stability, total active power to show overall generation and load balance, and voltage dip severity to describe the impact of disturbances. The problem is a multi-output regression because each output reflects a different part of the system behavior. Model performance is evaluated using prediction accuracy in both normal and disturbance conditions.

\subsection{Evaluation Objectives}

The following criteria are used to evaluate the surrogate models:

\begin{itemize}
\item \textbf{Prediction Accuracy:} This measures how well the models match the true system behavior, using root mean square error (RMSE), mean absolute error (MAE), and coefficient of determination ($R^2$).
\item \textbf{Robustness Under Uncertainty:} This shows how well the models handle real-world issues such as noise and communication delays.
\item \textbf{Computational Efficiency:} This measures how quickly the models run compared to EMT simulation and whether they are suitable for real-time use.
\end{itemize}

\section{SURROGATE MODELING METHODOLOGY}

\subsection{Overview}

\begin{figure}[ht]
\centering
\includegraphics[width=\columnwidth]{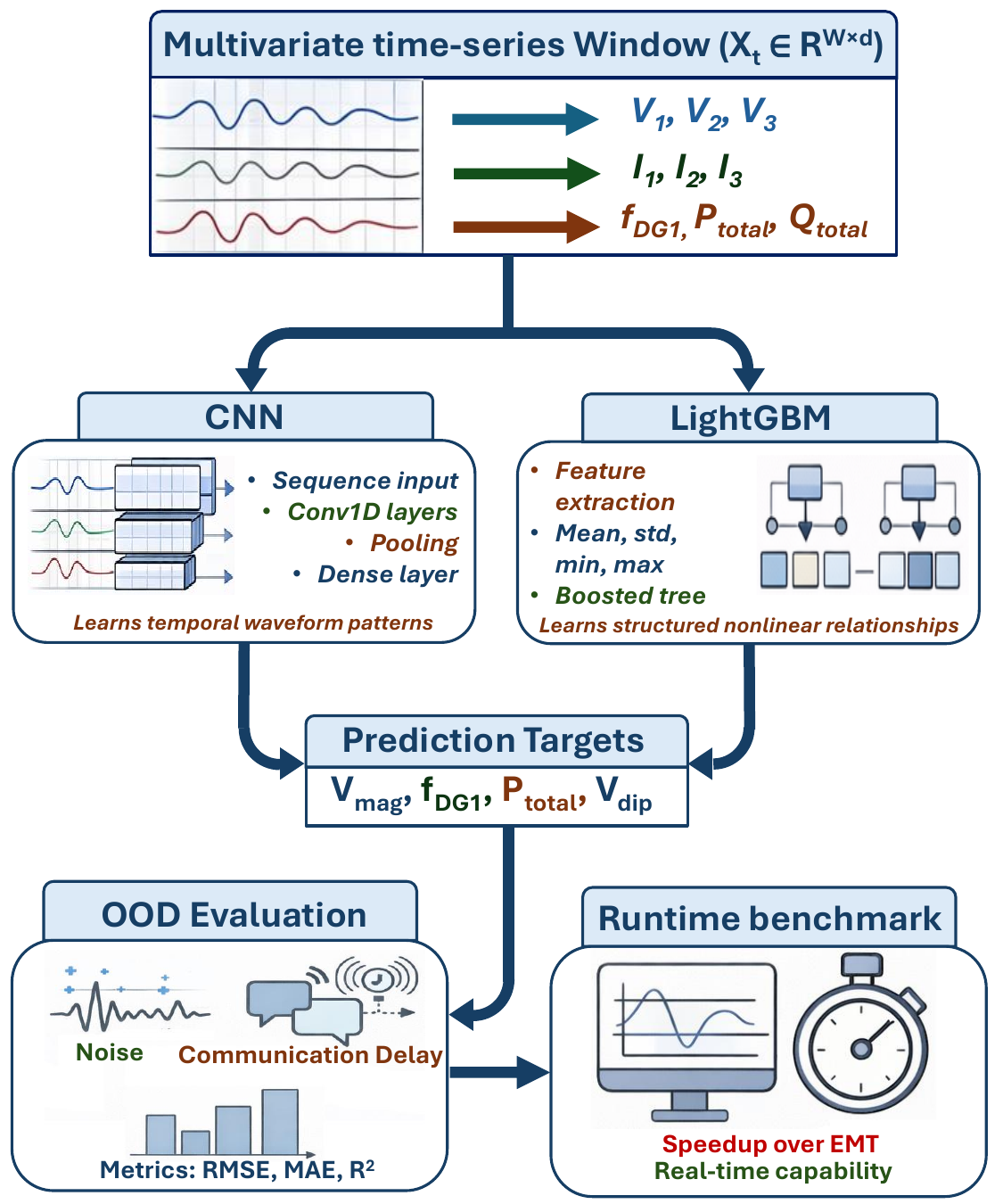}
\caption{Real-time surrogate modeling workflow using CNN and LightGBM. Multivariate time-series windows $X_t \in \mathbb{R}^{W \times d}$ are processed in parallel: CNN captures temporal waveform patterns, while LightGBM learns structured nonlinear relationships from statistical features. The models predict key variables $(V_{mag}, f_{DG1}, P_{total}, V_{dip})$, followed by OOD evaluation under noise and communication delay, and runtime benchmarking against EMT simulation to assess real-time capability.}
\label{fig:workflow}
\end{figure}

A surrogate modeling framework is presented in this section to approximate the dynamic behavior of the inverter-based microgrid \cite{mohammadi2024surrogate,marrel2024probabilistic,du2025development}. The system is built using two models: LightGBM and CNN. These models learn the relationship between time-windowed system measurements and key outputs. Data preprocessing, feature extraction, model training, and evaluation form the workflow. The key difference between the models lies in how they process time-based information. Figure~\ref{fig:workflow} illustrates the real-time surrogate modeling workflow. Two parallel approaches are used to process the multivariate time-series data: a CNN for temporal features and LightGBM for structured features. The results are then used to predict key system variables. The model is then evaluated under OOD conditions, and its runtime is compared with that of the EMT simulation to assess real-time use.

\subsection{Data Preprocessing and Feature Engineering}

Before training, the dataset is processed to make it suitable for modeling and to improve performance. All input features are normalized using standard scaling. Important system properties are captured by creating additional features from the raw data. The voltage magnitude is calculated from three-phase voltages as:

\[
V_{\text{mag}} = \sqrt{V_a^2 + V_b^2 + V_c^2}
\]

Total active and reactive power are also calculated by summing the outputs from all DG units. A sliding-window method with length $W$ is used to prepare the input data. For tree-based models, statistical features such as mean, standard deviation, minimum, maximum, and the most recent value are extracted from each window.

\subsection{LightGBM-Based Surrogate Model}

LightGBM is a machine learning method that uses gradient boosting to build decision trees one after another \cite{yang2024research,fu2025small}. It works well with structured data and can capture nonlinear relationships \cite{prina2024machine}. As a result, it forms an ensemble of decision trees in a stage-wise manner, which can be written as:
\[
F_M(x) = \sum_{m=1}^{M} \gamma_m h_m(x)
\]
where $M$ is the number of boosting stages, $h_m(x)$ is the $m$-th regression tree, and $\gamma_m$ is its corresponding weight.

The model minimizes a regularized objective:
\[
\mathcal{L} = \sum_{i=1}^{N} \ell(y_i, \hat{y}_i) + \sum_{m=1}^{M} \Omega(h_m)
\]

Here, $\ell(y_i,\hat{y}_i)$ is the prediction loss for sample $i$, and $\Omega(h_m)$ is the regularization term associated with the complexity of tree $h_m$. For each prediction task, a LightGBM model is trained using the extracted statistical features. The model learns from earlier mistakes to reduce prediction error. LightGBM can handle complex patterns, trains fast, and works well with engineered features. It also has low computational cost, which makes it suitable for real-time use. Accuracy is improved by training a separate model for each output variable. The main hyperparameters used for the LightGBM model are shown in Table~\ref{tab:model_config}.

\subsection{CNN-Based Surrogate Model}

A one-dimensional CNN is used to learn directly from raw time-series data \cite{shi2020convolutional,lee2023power,bogodorova2024fast}.
CNN models have also been shown to maintain strong performance under degraded or imperfect input conditions, which is important for real-world power system applications \cite{ogiesoba5264265robust}. It processes sliding-window sequences directly without manual feature design. The CNN is made up of several convolutional layers, activation functions, and pooling layers \cite{li2024power,zhang2025frequency}. The layers learn patterns from the input data, and fully connected layers are used to produce the final output. The convolution operation for a 1D CNN layer is defined as:
\[
z_i^{(l)} = \sigma \left( \sum_{k=0}^{K-1} w_k^{(l)} x_{i+k}^{(l-1)} + b^{(l)} \right)
\]
where $z_i^{(l)}$ is the output of layer $l$ at position $i$, $K$ is the convolution kernel size, $w_k^{(l)}$ are the filter weights, $x_{i+k}^{(l-1)}$ is the input from the previous layer, $b^{(l)}$ is the bias term, and $\sigma(\cdot)$ is the nonlinear activation function.

Once features are extracted by the convolutional layers, the final output is produced through fully connected layers and can be written as:

\[
\hat{y} = W_f z + b_f
\]

where $z$ is the learned feature representation from the final convolutional stage, $W_f$ is the fully connected weight matrix, and $b_f$ is the output bias vector. Using supervised learning, the model improves the match between predicted and true values. The CNN model learns patterns directly from raw data. It works well for transient behavior and waveform patterns and reduces the need for manual feature design. However, it usually requires more data and more computational resources \cite{lim2025power}. The architecture of the CNN model used in this study is summarized in Table~\ref{tab:model_config}.

\begin{table}[t]
\centering
\caption{Model Architecture and Hyperparameters Used in This Study}
\label{tab:model_config}
\begin{tabular}{p{0.48\columnwidth} p{0.48\columnwidth}}
\toprule
\textbf{CNN Architecture} & \textbf{LightGBM Hyperparameters} \\
\midrule
Conv1D (32): (100, 32) & Boosting: gbdt \\
Conv1D (64): (100, 64) & Learning rate: 0.05 \\
MaxPool: (50, 64) & Estimators: 300 \\
Conv1D (64): (50, 64) & Max depth: 6 \\
GlobalAvgPool: (64) & Num leaves: 31 \\
Dense (64) & Min child samples: 20 \\
Output (1) & Subsample: 0.8 \\
Total params: 85,541 & Colsample: 0.8 \\
Trainable: 28,513 & Random state: 42 \\
\bottomrule
\end{tabular}
\end{table}

\subsection{Training Strategy}

The models are trained on the dataset described earlier, which includes diverse operating conditions and disturbances.

\begin{itemize}
\item \textbf{Training set:} Used to learn the model from normal and disturbed data
\item \textbf{Validation set:} Used to tune the model and monitor performance
\item \textbf{Test set:} Includes noise and communication delay cases to test robustness \cite{aygul2024benchmark,oelhaf2025scoping}
\end{itemize}

To prevent overfitting, techniques such as early stopping and hyperparameter tuning are used. A separate model is used to train each output variable. Algorithm~\ref{alg:surrogate_workflow} summarizes the full workflow used for data preparation, surrogate model training, out-of-distribution testing, and runtime benchmarking.

\begin{algorithm}[t]
\small
\SetAlgoLined
\textbf{Input Data:} EMT scenario dataset $\mathcal{D}=\{d_1,d_2,\ldots,d_M\}$, window length $W$, stride $S$, input set $\mathcal{F}_{in}$, target set $\mathcal{F}_{out}$ \\
\textbf{Initialization:} train/validation/test scenario groups; CNN parameters $\theta_{\text{\sc{cnn}}}$; LightGBM parameters $\theta_{\text{\sc{lgbm}}}$ \\

\For{each scenario file $d_m \in \mathcal{D}$}{
    Load synchronized measurements $\{V_1,V_2,V_3,I_1,I_2,I_3,P_{DGk},Q_{DGk},f_{DGk}\}$ \\
    Compute derived variables:
    $V_{\text{mag}}=\sqrt{V_1^2+V_2^2+V_3^2}$,
    $P_{\text{total}}=\sum_{k=1}^{10} P_{DGk}$,
    $Q_{\text{total}}=\sum_{k=1}^{10} Q_{DGk}$ \\
    Append scenario label and scenario identifier \\
}

Concatenate all processed files to form $\mathcal{D}_{\text{all}}$ \\
Split $\mathcal{D}_{\text{all}}$ into train, validation, and OOD test subsets \\

\For{each subset $\mathcal{D}_s \in \{\mathcal{D}_{tr},\mathcal{D}_{val},\mathcal{D}_{te}\}$}{
    Construct sliding-window samples
    $X_t \in \mathbb{R}^{W \times |\mathcal{F}_{in}|}$ with stride $S$ \\
    Store raw sequence windows $X_t$ for CNN training \\
    Extract LightGBM feature vector
    $\phi(X_t)=\{\mu(X_t),\sigma(X_t),\min(X_t),\max(X_t),x_t^{\text{last}}\}$ \\
    Form target vector
    $y_t=\{V_{\text{mag}},\, f_{DG1},\, P_{\text{total}},\, V_{\text{dip}}\}$ \\
}

\For{each target variable $y \in \mathcal{F}_{out}$}{
    Train LightGBM regressor:
    $\hat{y}^{\text{\sc{lgbm}}}=f_{\text{\sc{lgbm}}}(\phi(X_t);\theta_{\text{\sc{lgbm}}})$ \\
    Normalize sequence inputs and target values \\
    Train CNN surrogate:
    $\hat{y}^{\text{\sc{cnn}}}=f_{\text{\sc{cnn}}}(X_t;\theta_{\text{\sc{cnn}}})$ \\
    Monitor validation loss and apply early stopping when needed \\
}

\For{each trained model}{
    Predict outputs on the OOD test set \\
    Compute performance metrics:
    $\mathrm{RMSE}$, $\mathrm{MAE}$, and $R^2$ \\
}

Measure surrogate inference time $T_{\text{\sc{cnn}}}$ and $T_{\text{\sc{lgbm}}}$ \\
Compare with EMT simulation runtime $T_{\text{\sc{emt}}}$ \\
Compute speedup:
$\Gamma_{\text{\sc{cnn}}}=T_{\text{\sc{emt}}}/T_{\text{\sc{cnn}}}$,
$\Gamma_{\text{\sc{lgbm}}}=T_{\text{\sc{emt}}}/T_{\text{\sc{lgbm}}}$ \\

\textbf{Output:} trained CNN and LightGBM surrogates, OOD prediction metrics, and runtime speedup results \\
\caption{Surrogate Modeling Workflow for Real-Time Microgrid Prediction}
\label{alg:surrogate_workflow}
\end{algorithm}

\subsection{Performance Evaluation Metrics}

The models are evaluated using RMSE, MAE, and $R^2$. These metrics show prediction accuracy, the average size of errors, and how well the model fits the data. Model speed is also evaluated by comparing prediction time with EMT simulation time to determine if real-time use is possible.

\subsection{Implementation Details}

The models were trained and tested using the same training, validation, and test datasets to ensure a fair comparison. They were created in Python, with LightGBM used for the tree-based model and a deep learning framework for the CNN. All experiments were run on a system with 50 GB RAM, an Intel Xeon CPU (8 cores, 2.0 GHz), and an NVIDIA Tesla T4 (16 GB) GPU for acceleration.

\section{RESULTS}

\subsection{Robustness Under Disturbances and Uncertainty}
\begin{figure*}[t]
\centering
\includegraphics[width=\textwidth]{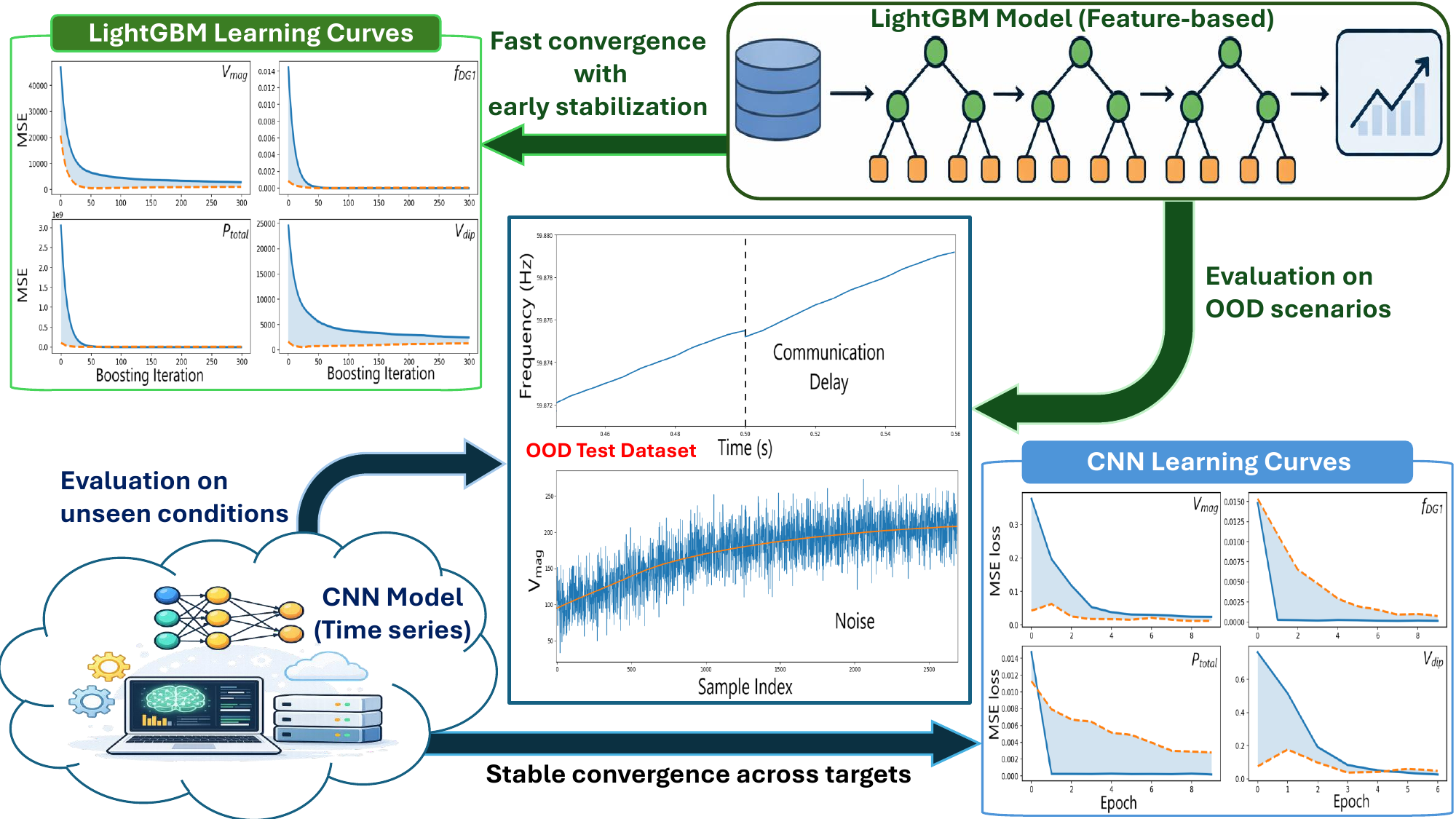}
\caption{Training and evaluation of CNN and LightGBM models under OOD conditions, such as noise and communication delay scenarios. The digital twin simulation is used to generate the training and validation data for both models. CNN converges steadily across all targets, while LightGBM converges faster and stabilizes early. Both models are tested on noise and communication delay scenarios.}
\label{fig:training_ood}
\end{figure*}

The surrogate models are evaluated on four important microgrid variables: voltage magnitude ($V_{\mathrm{mag}}$), system frequency ($f_{\mathrm{DG1}}$), total active power ($P_{\mathrm{total}}$), and voltage dip severity ($V_{\mathrm{dip}}$). The models are evaluated under challenging conditions, including noise and communication delays \cite{ogiesoba2026high, aygul2024benchmark}. 
Fig.~\ref{fig:training_ood} shows the training behavior and OOD evaluation of both models. The learning curves of the training loss (MSE) show that the CNN converges steadily to capture time-dependent patterns across the target variables $V_{\mathrm{mag}}, f_{\mathrm{DG1}}, P_{\mathrm{total}}, and V_{\mathrm{dip}}$). LightGBM converges faster and stabilizes early to learn feature-based relationships. Both models maintained fairly stable prediction performance under OOD scenarios with disturbances, which indicates good generalization to unseen operating conditions. This means they were able to predict microgrid variables reliably even when the input data was affected by noise and communication delays.

\subsection{Prediction Performance Across Target Variables}

\begin{figure}[ht]
\centering
\includegraphics[width=\columnwidth]{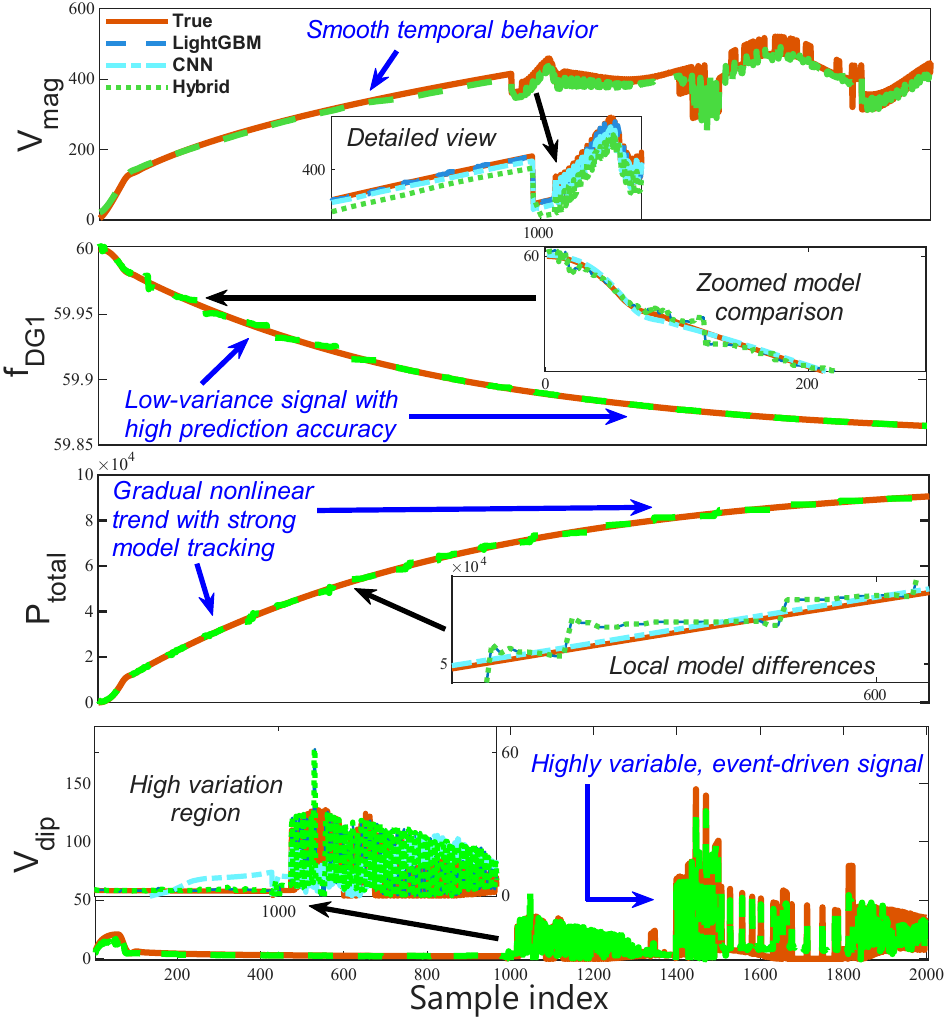}
\caption{Comparison of model predictions and ground truth under unseen conditions for key microgrid variables: voltage magnitude ($V_{\mathrm{mag}}$), frequency ($f_{\mathrm{DG1}}$), total active power ($P_{\mathrm{total}}$), and voltage dip ($V_{\mathrm{dip}}$). Results from CNN, LightGBM, and the hybrid model are shown. Insets show small local differences between the models. Smooth variables ($V_{\mathrm{mag}}$, $P_{\mathrm{total}}$) and frequency are predicted accurately, while $V_{\mathrm{dip}}$ shows highly variable behavior with more visible deviations. The hybrid model provides consistent performance across both smooth and dynamic signals.}
\label{fig:OOD_Model}
\end{figure}

Fig.~\ref{fig:OOD_Model} presents the prediction results of CNN, LightGBM, and the hybrid model for the selected variables. For slowly varying signals such as voltage magnitude ($V_{\mathrm{mag}}$) and total active power ($P_{\mathrm{total}}$), all models track the true values well. The hybrid model shows better alignment during small local changes. Frequency ($f_{\mathrm{DG1}}$) is predicted with very high accuracy ($R^2 > 0.99$), mainly because it changes smoothly and has low variability. In contrast, voltage dip ($V_{\mathrm{dip}}$) is more difficult to model because it is highly variable and driven by sudden events. As a result, the differences between the models become more clearly visible. The hybrid model maintains stable performance by combining the strengths of both CNN and LightGBM, showing the importance of hybrid approaches for complex microgrid dynamics.

\subsection{Residual Analysis and Error Distribution}

\begin{figure}[ht]
\centering
\includegraphics[width=\columnwidth]{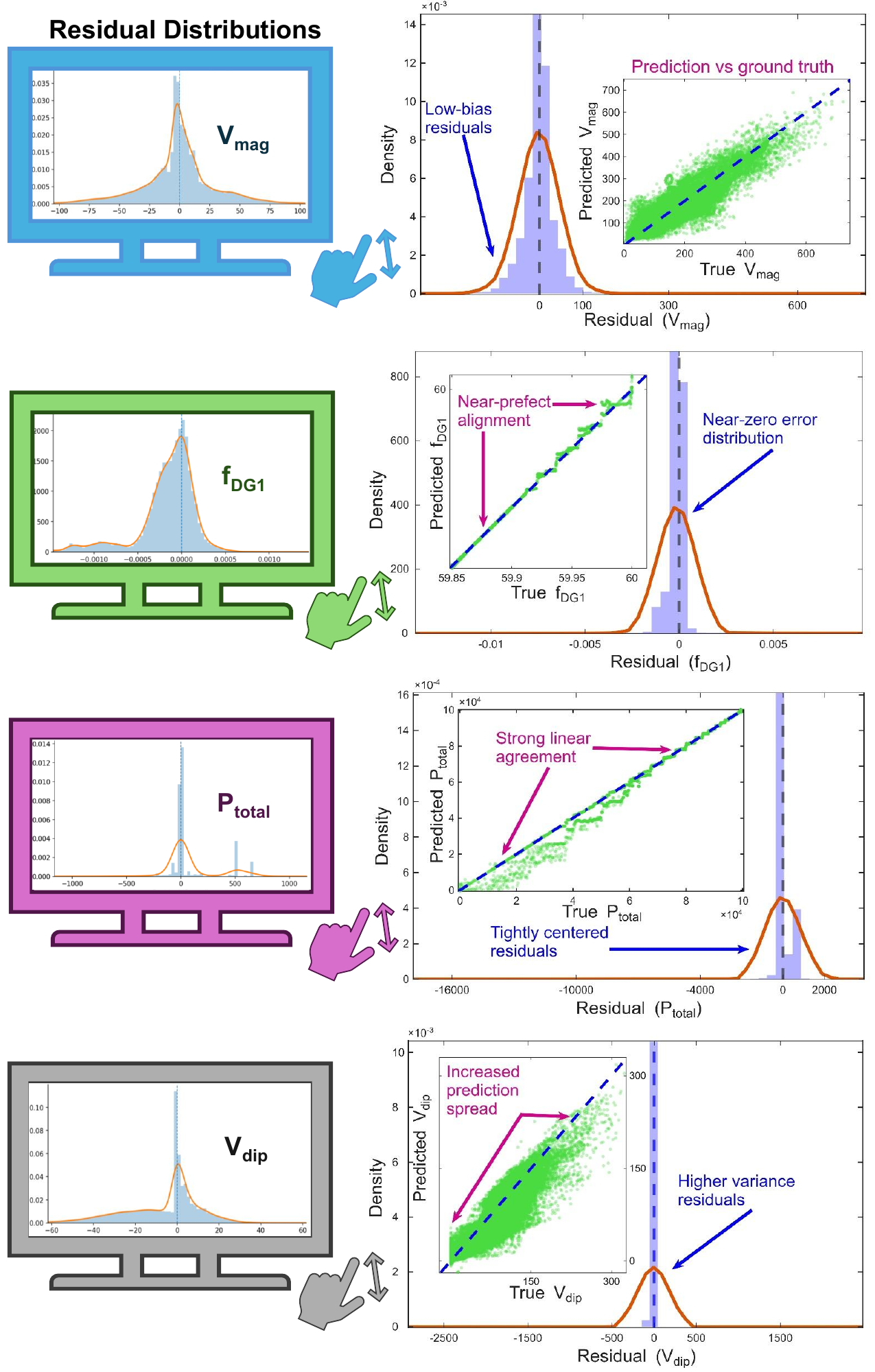}
\caption{Residual distributions and prediction accuracy for key microgrid variables: voltage magnitude ($V_{\mathrm{mag}}$), frequency ($f_{\mathrm{DG1}}$), total active power ($P_{\mathrm{total}}$), and voltage dip ($V_{\mathrm{dip}}$). Histograms display the residual distributions along with kernel density estimates, while the insets show predicted versus true values. Most variables have residuals that are tightly centered around zero with low bias, and they show strong linear agreement, especially for $f_{\mathrm{DG1}}$ and $P_{\mathrm{total}}$. In contrast, $V_{\mathrm{dip}}$ shows higher variance and increased prediction spread due to its highly dynamic behavior.}
\label{fig:residuals}
\end{figure}

Residual distributions and prediction alignment are used to better evaluate model accuracy. Fig.~\ref{fig:residuals} shows that the residuals for $V_{\mathrm{mag}}$, $f_{\mathrm{DG1}}$, and $P_{\mathrm{total}}$ are tightly centered around zero, which indicates low bias and high accuracy. The inset plots further show that the predicted values closely match the true values. In contrast, $V_{\mathrm{dip}}$ has a wider residual distribution and more variation in its predictions, showing that it is harder to model because of its highly dynamic behavior. Despite this difficulty, the models still achieve acceptable accuracy for both smooth and complex variables. For time-dependent signals such as $V_{\mathrm{mag}}$, CNN performs better, while LightGBM is more effective for derived variables such as $V_{\mathrm{dip}}$. This shows the benefit of combining temporal and feature-based learning.

\subsection{Computational Efficiency and Real-Time Feasibility}

Computational speed is essential for real-time microgrid applications. 
The execution time of each surrogate model is compared with the EMT-based Simulink simulation. The real-time (RT) ratio is the ratio of simulated time to wall-clock execution time. Table~\ref{tab:runtime} shows that all surrogate models significantly reduce computation time compared to the EMT simulation. LightGBM achieves the highest speedup ($>1000\times$) and runs faster than real time (RT ratio $>1$). The hybrid model achieves over $500\times$ speedup and runs close to real-time. Although the CNN model is accurate, it does not reach real-time performance because of its higher computational cost.

\begin{table}[t]
\centering
\caption{Runtime and Real-Time Performance Comparison}
\label{tab:runtime}
\begin{tabular}{lcccc}
\toprule
\textbf{Method} & \textbf{Sim (s)} & \textbf{Time (s)} & \textbf{Speedup} & \textbf{RT Ratio} \\
\midrule
Simulink & 1.00 & 941.16 & 1.00 & 0.001 \\
LightGBM & 1.00 & 0.89 & 1053.58 & 1.12 \\
CNN & 1.00 & 5.09 & 185.04 & 0.20 \\
Hybrid & 1.00 & 1.80 & 522.65 & 0.56 \\
\bottomrule
\end{tabular}
\end{table}

\begin{figure}[ht]
\centering
\includegraphics[width=\columnwidth]{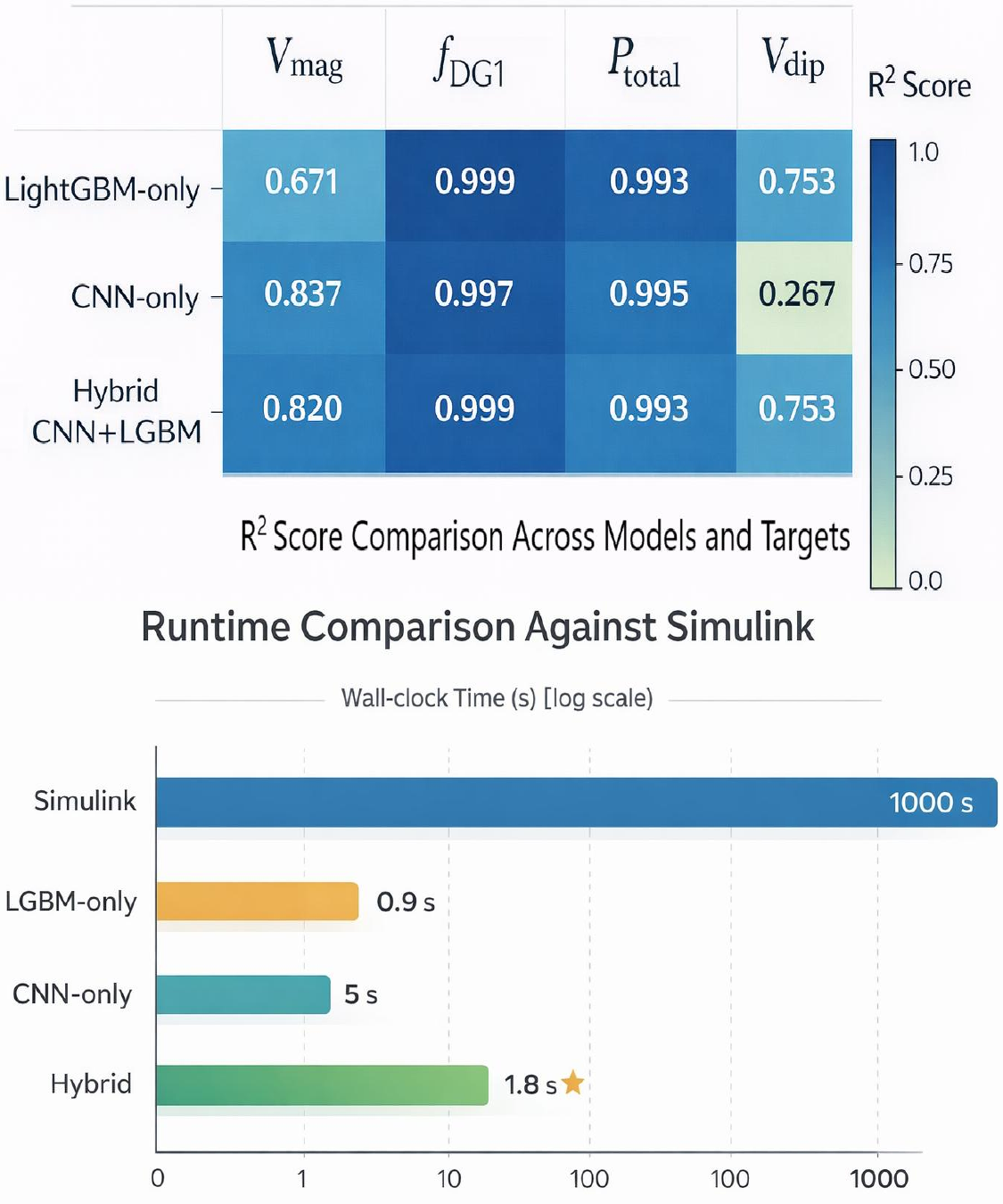}
\caption{Combined evaluation of prediction accuracy and computational efficiency. The top panel shows $R^2$ scores for LightGBM, CNN, and the hybrid model across key variables: $V_{\mathrm{mag}}$, $f_{\mathrm{DG1}}$, $P_{\mathrm{total}}$, and $V_{\mathrm{dip}}$. The bottom panel presents runtime on a logarithmic scale relative to the EMT simulation.}
\label{fig:combined_perf_runtime}
\end{figure}

Fig.~\ref{fig:combined_perf_runtime} summarizes both accuracy and efficiency. Model performance depends on the type of variable being predicted. CNN achieves the highest accuracy for voltage magnitude ($R^2 = 0.837$) because it effectively captures time-based patterns \cite{shi2020convolutional,lee2023power,li2024power}. For frequency prediction, both models are highly accurate, but LightGBM performs slightly better because the signal is smooth and predictable \cite{fu2025small,zhang2025frequency,yang2024research}. In the same way, both models perform well for total active power, with CNN having a slight advantage. In contrast, LightGBM performs much better for voltage dip ($R^2 = 0.753$ vs.\ 0.267) because it is a derived variable that is easier to capture using feature-based relationships \cite{yang2024research,fu2025small}. While all models achieve high accuracy for $f_{\mathrm{DG1}}$ and $P_{\mathrm{total}}$, performance varies for $V_{\mathrm{mag}}$ and $V_{\mathrm{dip}}$. By combining the outputs of CNN and LightGBM, the hybrid model captures the strengths of both across different variables. This shows that temporal learning (CNN) and feature-based learning (LightGBM) work well together and support the use of hybrid models for different types of microgrid dynamics. Overall, the hybrid model maintains consistently strong performance across all variables. In addition, surrogate models reduce runtime by orders of magnitude, with the hybrid approach providing a strong balance between accuracy and computational efficiency, making it suitable for real-time applications.

\section{Discussion}

The results give useful insights into how different surrogate models behave when applied to inverter-based microgrids. Both CNN and LightGBM give high accuracy, but their performance is not the same for all tasks. This shows that model selection should depend on the type of variable being predicted, rather than a one-size-fits-all approach. One key finding is that model performance depends on the type of variable. Variables like voltage change over time due to inverter control and fast system responses. These signals depend on past values, so CNN works well because it can learn patterns from time-series data \cite{shi2020convolutional,lee2023power,bogodorova2024fast}. On the other hand, frequency is a smooth and low-variance signal, while voltage dip is a derived and event-driven variable. These variables depend more on overall system behavior than on detailed time patterns. Frequency changes smoothly and is easier to predict using statistical features \cite{zhang2025frequency}. Voltage dip is derived from the signal during disturbances, so LightGBM performs better, as it can learn from engineered features without relying on full time-based modeling \cite{yang2024research,fu2025small}. This shows that no single model can capture all types of system behavior. Different models work better for different types of variables. Another important point is the role of feature engineering. LightGBM performs well for voltage dip because it uses derived features. Voltage dip is not directly seen in raw data but is calculated from it. As a result, CNN is less effective because it relies on raw time-series input and may not fully capture these derived features. This means feature engineering is still important. Even though deep learning reduces the need for manual features, it may not work well when the target is not directly present in the data. Better results can be achieved by combining domain knowledge with data-driven methods \cite{ellinas2025physics}. The models are tested under noise and communication delay conditions to represent real-world microgrid data. The results show that the models perform well in these situations, meaning they can handle these variations \cite{aygul2024benchmark,barreto2025cyber}. Both models remain stable for frequency prediction, while LightGBM performs better for voltage dip. CNN performs well for voltage but is less effective for derived variables such as voltage dip. This makes it clear that models should be tested under uncertain conditions. From a practical point of view, speed is very important. LightGBM is very fast and requires less computation, making it suitable for real-time applications and edge devices \cite{cheng2024machine,du2025development}. CNN needs more computation but gives better results for time-based signals. The models are much faster than EMT simulations, which makes them useful for real-time monitoring, fast analysis, and decision-making. However, selecting a model should take into account both accuracy and speed. The models help operators respond faster to disturbances by quickly predicting how the system behaves. This makes them useful for real-time microgrid operation. For example, LightGBM can quickly estimate frequency and disturbance levels, while CNN can be used for more detailed voltage analysis. This shows that better results can be achieved by using multiple models, where each model handles different variables \cite{islam2025machine}. Some limitations exist in this study. Because the models are trained on one microgrid, they may need more data to perform well on other systems. The current approach uses single-step prediction, so it may not reflect long-term system behavior. Improving the models with multi-step prediction, sequence learning, and advanced models like transformers will be the next step \cite{lim2025power,islam2025machine}. It will also involve combining them with physical knowledge and control systems to improve accuracy and support real-time operation.

\section{Conclusion}
This paper presents a data-driven surrogate modeling framework for real-time prediction of inverter-based microgrid behavior using CNN and LightGBM models. The models are trained on a high-quality EMT digital twin dataset that includes different operating conditions and disturbances. The type of variable being predicted affects model performance. CNN performs better for time-dependent variables such as voltage magnitude. LightGBM performs slightly better for smooth variables, such as frequency, and also for derived variables such as voltage dip. Therefore, no single model works best for all variables, and model selection should be based on the physical characteristics of each variable. The surrogate models are both accurate and much faster than the EMT simulation. LightGBM runs faster than real time with very high speedup, whereas the hybrid CNN+LightGBM model maintains consistent performance while operating close to real time. This makes them suitable for real-time monitoring, fault analysis, and decision-making in microgrids. The models perform well under real-world conditions like noise and communication delays, making them suitable for practical use. To improve accuracy and real-time performance, future work will focus on multi-step prediction, use advanced models like transformers, and combine data-driven methods with physical knowledge.

\bibliographystyle{IEEEtran}   
\bibliography{references}      

@article{ogiesoba2026high,
  title={High-Fidelity Digital Twin Dataset Generation for Inverter-Based Microgrids Under Multi-Scenario Disturbances},
  author={Ogiesoba-Eguakun, Osasumwen Cedric and Ashenayi, Kaveh and Rath, Suman},
  journal={arXiv preprint arXiv:2603.10262},
  year={2026}
}

@article{aghazadeh2024digital,
  title={Digital Twins of smart energy systems: a systematic literature review on enablers, design, management and computational challenges},
  author={Aghazadeh Ardebili, Ali and Zappatore, Marco and Ramadan, Amro Issam Hamed Attia and Longo, Antonella and Ficarella, Antonio},
  journal={Energy Informatics},
  volume={7},
  number={1},
  pages={94},
  year={2024},
  publisher={Springer}
}

@article{aghdam2025navigating,
  title={Navigating the digital landscape: A review of digitalization in smart grids with renewable energy sources},
  author={Aghdam, Farid Hamzeh and Zavodovski, Aleksandr and Rasti, Mehdi and Pongracz, Eva},
  journal={Journal of Renewable and Sustainable Energy},
  volume={17},
  number={5},
  year={2025},
  publisher={AIP Publishing}
}

@article{aravena2025open,
  title={Open Power System Datasets and Open Simulation Engines: A Survey Towards Machine Learning Applications},
  author={Aravena, Ignacio and Sun, Chih-Che and Shi, Ranyu and Majumder, Subir and Yan, Weihang and Joo, Jhi-Young and Xie, Le and Wang, Jiyu},
  journal={IEEE Open Access Journal of Power and Energy},
  year={2025},
  publisher={IEEE}
}

@article{barreto2025cyber,
  title={Cyber-Physical Power System Digital Twins—A Study on the State of the Art},
  author={Barreto, Nathan Elias Maruch and Aoki, Alexandre Rasi},
  journal={Energies},
  volume={18},
  number={22},
  pages={5960},
  year={2025},
  publisher={MDPI}
}

@article{kabir2024digital,
  title={Digital twins for IoT-driven energy systems: A survey},
  author={Kabir, Md Rafiul and Halder, Dipal and Ray, Sandip},
  journal={IEEE Access},
  volume={12},
  pages={177123--177143},
  year={2024},
  publisher={IEEE}
}

@article{jiang2024digital,
  title={Digital twin of microgrid for predictive power control to buildings},
  author={Jiang, Hao and Tjandra, Rudy and Soh, Chew Beng and Cao, Shuyu and Soh, Donny Cheng Lock and Tan, Kuan Tak and Tseng, King Jet and Krishnan, Sivaneasan Bala},
  journal={Sustainability},
  volume={16},
  number={2},
  pages={482},
  year={2024},
  publisher={MDPI}
}

@article{mbasso2025digital,
  title={Digital twins in renewable energy systems: A comprehensive review of concepts, applications, and future directions},
  author={Mbasso, Wulfran Fendzi and Harrison, Ambe and Dagal, Idriss and Jangir, Pradeep and Khishe, Mohammad and Kotb, Hossam and Shaikh, Muhammad Suhail and Smerat, Aseel and Donfack, Emmanuel Fendzi and Kumar, Raman},
  journal={Energy Strategy Reviews},
  volume={61},
  pages={101814},
  year={2025},
  publisher={Elsevier}
}

@article{thwe2025digital,
  title={Digital twins for power systems: Review of current practices, requirements, enabling technologies, data federation and challenges},
  author={Thwe, May Myat and {\c{S}}tefanov, Alexandru and Rajkumar, Vetrivel Subramaniam and Palensky, Peter},
  journal={IEEE Access},
  year={2025},
  publisher={IEEE}
}

@article{samal2025review,
  title={A review on microgrid control: Conventional, advanced and intelligent control approaches},
  author={Samal, Kalpana Bijayeeni and Mahapatra, Mitali and Pati, Swagat and Debnath, Manoj Kumar},
  journal={Unconventional Resources},
  pages={100297},
  year={2025},
  publisher={Elsevier}
}

@article{islam2025machine,
  title={Machine learning for power system stability and control},
  author={Islam, Rakibul and Rivin, Mir Araf Hossain and Sultana, Sharmin and Asif, MD Amaddus Bepary and Mohammad, Mahathir and Rahaman, Mustafizur},
  journal={Results in engineering},
  volume={26},
  pages={105355},
  year={2025},
  publisher={Elsevier}
}

@article{lim2025power,
  title={Power System Decision Making in the Age of Deep Learning: A Comprehensive Review.},
  author={Lim, Yeji and Son, Minjae and Park, Kyungnam and Kim, Minsoo and Song, Keunju and Lee, Haejoong and Kim, Hongseok},
  journal={Energies (19961073)},
  volume={18},
  number={18},
  year={2025}
}

@article{r2024machine,
  title={Machine learning-based energy management and power forecasting in grid-connected microgrids with multiple distributed energy sources},
  author={R. Singh, Arvind and Kumar, R Seshu and Bajaj, Mohit and Khadse, Chetan B and Zaitsev, Ievgen},
  journal={Scientific Reports},
  volume={14},
  number={1},
  pages={19207},
  year={2024},
  publisher={Nature Publishing Group UK London}
}

@article{cheng2024machine,
  title={Machine-learning-reinforced massively parallel transient simulation for large-scale renewable-energy-integrated power systems},
  author={Cheng, Tianshi and Chen, Ruogu and Lin, Ning and Liang, Tian and Dinavahi, Venkata},
  journal={IEEE Transactions on Power Systems},
  volume={40},
  number={1},
  pages={970--981},
  year={2024},
  publisher={IEEE}
}

@article{mohammadi2024surrogate,
  title={Surrogate modeling for solving OPF: A review},
  author={Mohammadi, Sina and Bui, Van-Hai and Su, Wencong and Wang, Bin},
  journal={Sustainability},
  volume={16},
  number={22},
  pages={9851},
  year={2024},
  publisher={MDPI}
}

@article{marrel2024probabilistic,
  title={Probabilistic surrogate modeling by Gaussian process: A review on recent insights in estimation and validation},
  author={Marrel, Amandine and Iooss, Bertrand},
  journal={Reliability Engineering \& System Safety},
  volume={247},
  pages={110094},
  year={2024},
  publisher={Elsevier}
}

@article{prina2024machine,
  title={Machine learning as a surrogate model for EnergyPLAN: Speeding up energy system optimization at the country level},
  author={Prina, Matteo Giacomo and Dallapiccola, Mattia and Moser, David and Sparber, Wolfram},
  journal={Energy},
  volume={307},
  pages={132735},
  year={2024},
  publisher={Elsevier}
}

@article{du2025development,
  title={Development of a robust data-driven surrogate model to improve energy flexibility of an integrated district heating system with a thermal storage system},
  author={Du, Han and Zhou, Xinlei and Nord, Natasa and Carden, Yale and Cui, Ping and Ma, Zhenjun},
  journal={Energy},
  pages={138398},
  year={2025},
  publisher={Elsevier}
}

@article{ellinas2025physics,
  title={Physics-informed machine learning for power system dynamics: A framework incorporating trustworthiness},
  author={Ellinas, Petros and Karampinis, Ioannis and Nadal, Ignasi Ventura and Nellikkath, Rahul and Vorwerk, Johanna and Chatzivasileiadis, Spyros},
  journal={Sustainable Energy, Grids and Networks},
  pages={101818},
  year={2025},
  publisher={Elsevier}
}

@article{oelhaf2025scoping,
  title={A scoping review of machine learning applications in power system protection and disturbance management},
  author={Oelhaf, Julian and Kordowich, Georg and Pashaei, Mehran and Bergler, Christian and Maier, Andreas and J{\"a}ger, Johann and Bayer, Siming},
  journal={International Journal of Electrical Power \& Energy Systems},
  volume={172},
  pages={111257},
  year={2025},
  publisher={Elsevier}
}

@article{von2023power,
  title={Power hardware-in-the-loop (PHIL): A review to advance smart inverter-based grid-edge solutions},
  author={von Jouanne, Annette and Agamloh, Emmanuel and Yokochi, Alex},
  journal={Energies},
  volume={16},
  number={2},
  pages={916},
  year={2023},
  publisher={MDPI}
}

@article{zhang2021critical,
  title={A critical review of data-driven transient stability assessment of power systems: principles, prospects and challenges},
  author={Zhang, Shitu and Zhu, Zhixun and Li, Yang},
  journal={Energies},
  volume={14},
  number={21},
  pages={7238},
  year={2021},
  publisher={MDPI}
}

@article{shi2020convolutional,
  title={Convolutional neural network-based power system transient stability assessment and instability mode prediction},
  author={Shi, Zhongtuo and Yao, Wei and Zeng, Lingkang and Wen, Jianfeng and Fang, Jiakun and Ai, Xiaomeng and Wen, Jinyu},
  journal={Applied Energy},
  volume={263},
  pages={114586},
  year={2020},
  publisher={Elsevier}
}

@article{pournabi2022power,
  title={Power system transient security assessment based on deep learning considering partial observability},
  author={Pournabi, Mehrdad and Mohammadi, Mohammad and Afrasiabi, Shahabodin and Setoodeh, Peyman},
  journal={Electric Power Systems Research},
  volume={205},
  pages={107736},
  year={2022},
  publisher={Elsevier}
}

@article{lee2023power,
  title={Power system transient stability assessment using convolutional neural network and saliency map},
  author={Lee, Heungseok and Kim, Jongju and Park, June Ho and Chung, Sang-Hwa},
  journal={Energies},
  volume={16},
  number={23},
  pages={7743},
  year={2023},
  publisher={MDPI}
}

@article{bogodorova2024fast,
  title={Fast small signal stability assessment using deep convolutional neural networks},
  author={Bogodorova, Tetiana and Osipov, Denis and Vanfretti, Luigi},
  journal={Electric Power Systems Research},
  volume={235},
  pages={110853},
  year={2024},
  publisher={Elsevier}
}

@article{li2024power,
  title={Power system transient voltage vulnerability assessment based on knowledge visualization of CNN},
  author={Li, Zhendong and Yan, Jiongcheng and Liu, Yutian and Liu, Weipeng and Li, Li and Qu, Hanbing},
  journal={International Journal of Electrical Power \& Energy Systems},
  volume={155},
  pages={109576},
  year={2024},
  publisher={Elsevier}
}

@article{aygul2024benchmark,
  title={Benchmark of machine learning algorithms on transient stability prediction in renewable rich power grids under cyber-attacks},
  author={Aygul, Kemal and Mohammadpourfard, Mostafa and Kesici, Mert and Kucuktezcan, Fatih and Genc, Istemihan},
  journal={Internet of Things},
  volume={25},
  pages={101012},
  year={2024},
  publisher={Elsevier}
}

@article{yang2024research,
  title={Research on power system small signal stability analysis and correction based on LightGBM algorithm},
  author={Yang, Yude and Wang, Yijun and Zhang, Xiu},
  journal={Electrical Engineering},
  volume={106},
  number={4},
  pages={4469--4486},
  year={2024},
  publisher={Springer}
}

@article{fu2025small,
  title={Small signal stability prediction and correction control algorithm for wind power systems based on LightGBM},
  author={Fu, Yang and Guosong, Wang and Yi, Xu and Qinfeng, Ma and Su, An and Junquan, Chen},
  journal={Frontiers in Energy Research},
  volume={13},
  pages={1725371},
  year={2025},
  publisher={Frontiers Media SA}
}

@article{zhang2025frequency,
  title={On frequency stability assessment of power systems using convolutional neural network-Informer and deep learning important features},
  author={Zhang, Yihao and Han, Song and Rong, Na and Liu, Wensheng},
  journal={Sustainable Energy, Grids and Networks},
  pages={102056},
  year={2025},
  publisher={Elsevier}
}

@article{ogiesoba5264265robust,
  title={Robust Cnn-Based Multi-Class Object Recognition High Accuracy on Blurred Images for Real-World Situational Awareness Systems},
  author={Ogiesoba-Eguakun, Osasumwen Cedric and Idonor, Chidinma Ezekiel},
  journal={Available at SSRN 5264265},
  year={2025}
}

\vfill

\end{document}